# Energy Minimized Federated Fog Computing over Passive Optical Networks

Abdullah M. Alqahtani, Barzan Yosuf, Sanaa H. Mohamed, Taisir E.H. El-Gorashi, and Jaafar M.H. Elmirghani, Fellow, IEEE
School of Electronic and Electrical Engineering, University of Leeds, LS2 9JT, United Kingdom

*Abstract*— The rapid growth of time-sensitive applications and services has driven enhancements to computing infrastructures. The main challenge that needs addressing for these applications is the optimal placement of the end-users' demands to reduce the total power consumption and delay. One of the widely adopted paradigms to address such a challenge is fog computing. Placing fog units close to end-users at the edge of the network can help mitigate some of the latency and energy efficiency issues. Compared to the traditional hyperscale cloud data centres, fog computing units are constrained by computational power, hence, the capacity of fog units plays a critical role in meeting the stringent demands of the end-users due to intensive processing workloads. In this paper, we aim to optimize the placement of virtual machines (VMs) demands originating from end-users in a fog computing setting by formulating a Mixed Integer Linear Programming (MILP) model to minimize the total power consumption through the use of a federated architecture made up of multiple distributed fog cells. The obtained results show an increase in processing capacity in the fog layer and a reduction in the power consumption by up to 26% compared to the Non-Federated fogs network.

I. INTRODUCTION

The growth in next generation applications is accelerated by the Internet of Things (IoT), smart cities, healthcare, and smart grids to name a few. This is expected to result in the generation of enormous volumes of data that is usually transported over multiple network domain towards remote cloud data centers for processing purposes. Processing all that data by the cloud will cause further bottlenecks in the already congested core network and hence, this will have a detrimental impact on latency and energy efficiency. Therefore, through the concept of fog computing, computational resources are placed close to the end-users, at the edge of the network, in order to extend the capabilities of cloud data centers. Fog computing units could be any device that can feature compute and processing capability such as routers, switches, accesses points or small racks of servers [1]–[8]. However, the processing capacity of fog units is limited compared to the cloud, often resulting in congestion or virtual machine (VM) blockage when there is a large amount of data for processing. Therefore, the capacity of fog computing units should be addressed to ensure that they are well-suited to the demands of time sensitive applications [9]–[11]. The authors in [9] optimized the service allocation problem using an integer programming model in a cloud-fog architecture. The goal of the work is to minimize the latency experienced by IoT services while meeting resource constraints. Two modes of service allocation are considered which the authors refer to as serial and parallel allocation. With the former, higher delays are experienced by the services, whilst with the latter approach lower service delays are observed. The study in [10] reports on the work of a consortium called RECAP that aims to advance cloud and edge technologies to develop mechanisms for reliable capacity provisioning as well as making application placement and infrastructure orchestration autonomous, predictable and optimized. This automation is achieved by intelligent profiling of workloads. Passive Optical Networks (PONs) are widely utilized as an optimal technology in the access part of the network as well as the cloud domain due to their energy efficiency and high bandwidths that are particularly suited to high bit-rate applications such as streaming, VoD and cloud gaming, etc. [12]–[15]. The authors in [15] assessed wired and wireless network infrastructures in terms of their applicability to real-time applications. The study in [15] shows that PON is a power efficient access technology due to passive nodes, with the additional advantages of abundant bandwidth.

The work in [16] studies the performance of multiple cooperative fog servers in a wireless network by evaluating the users' quality of experience (QoE) and fog nodes' power efficiency. The trade-off between these two metrics are discussed for a single-node fog computing network as well as extending the study to look at fog node cooperation. The study comprises of a workload allocation problem that is solved through a distributed optimization algorithm, which achieves comparable results to the results obtained by their global optimization model. Moreover, the authors in [17] focus on minimizing the total service latency of multiple users with homogenous tasks. Their fog radio access network (F-RAN) model utilizes the existing wireless infrastructure such as small cells and near-range communication links at the edge layer. Different from the aforementioned studies, in this paper, we optimize and evaluate a practical energy efficient fog computing architecture through a federated fog approach in which neighbouring fog cells collaborate to complete a given service. We benefit from our previous work in energy efficiency that tackled areas such as distributed processing in the IoT/Fog [18]–[21], green core and data centre (DC) networks [22]–[31],[32]–[37], network virtualization and service embedding in core and IoT networks [38]–[41] and machine learning and network optimization for healthcare systems [42]–[45] and network coding in the core network [46], [47].

The reminder of this paper is organized as follows: Section 2 explains the proposed architecture and the optimization model.

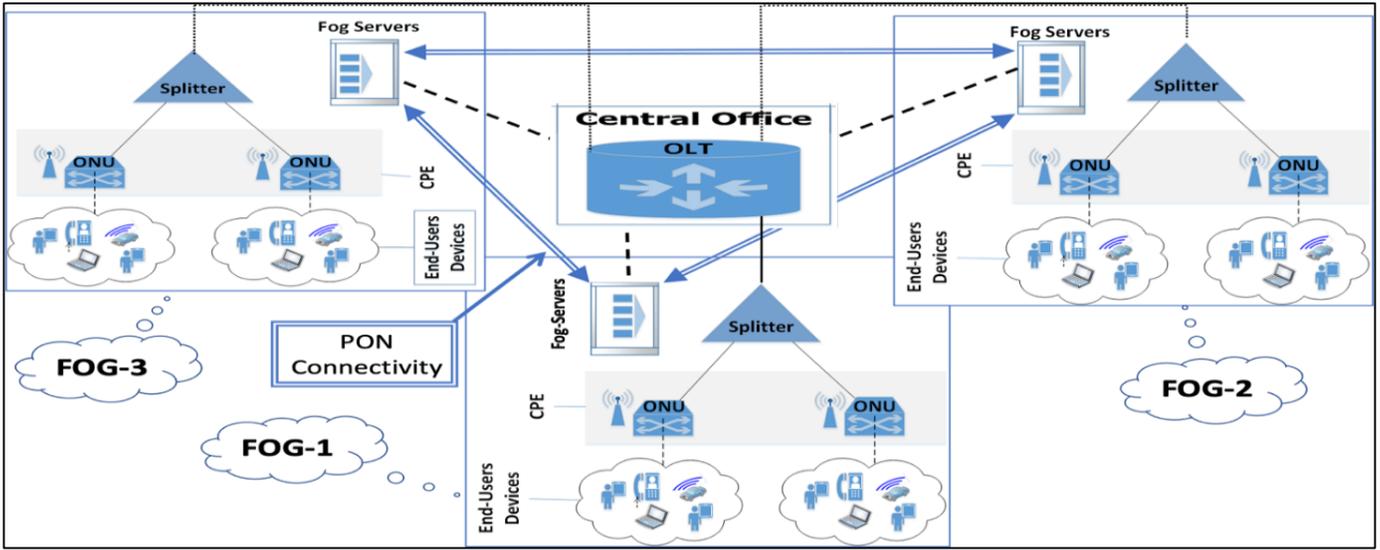

Figure 1 The Federated Fog Architecture over a Passive Optical Network (PON).

Section 3 presents and discusses the optimization results. Finally, Section 4 concludes the paper and outlines avenues for future research.

## II. THE PROPOSED FOG-BASED PON ARCHITECTURE

Time-sensitive applications should be processed in the nearest computing units close to the end-devices in order to meet the low-latency requirements of these applications. However, due to the size of the collected data at the edge of the network, adaptions are needed in the computing architectures to tackle their limited capacity so that large amounts of data can be processed closer to the data sources. Therefore, the proposed fog based PON architecture provides connectivity between three (or more) fog computing cells. Hence, in a federated fog network, multiple fog cells are able to collaborate to host the user demands and subsequently reduce the total power consumption compared to a non-federated network with non-collaborative fog cells. Moreover, the proposed architecture, as shown in Fig. 1, is comprised of a networking layer and a processing layer.

### A. The Networking Layer

The networking layer is responsible for aggregating the data from the end-user's equipment and forwarding it to the processing units. In the proposed architecture, the networking layer is composed of energy efficient devices within the PON architecture that are comprised of the Optical Networking Units (ONU), that represent the Customer Premises Equipment (CPE). These are in charge of collecting data from end-users. In addition, an Optical Line Terminal (OLT) at the Central Office, is in charge of collecting data from the ONUs, and a passive splitter between the ONUs and OLT provides a point-to-multipoint fibre optical network.

### B. The Processing Layer

The processing layer is responsible for hosting the virtual machines requests (VMs) generated from the end-user's devices and processing the data in efficient ways to meet the delay sensitive applications requirements. In the proposed architecture, the processing layer is comprised of fog servers, where each fog cell has multiple fog servers. Each server is equipped with an ONU device.

### C. MILP Optmization Model

The proposed MILP model aims to optimize the placement of VM demands and augment the capacity of fog cells in the proposed architecture with the objective of minimizing the total power consumption jointly with minimizing the number of blocked VMs. The model optimizes the PON connectivity among the neighbouring fog computing cells so that over-demanding VM requests on one of the fog cells can be processed in the closet connected available fog cell. Note that in the proposed federated fog architecture, there is PON connectivity between the fog cells to facilitate "borrowing" of data processing capabilities and all of the fog cells are

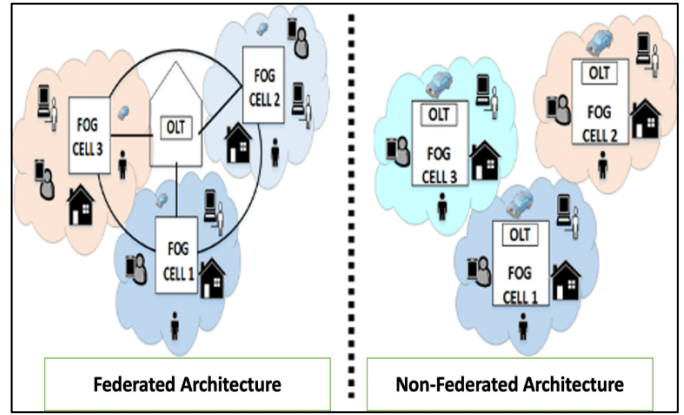

Figure 2 PON-Based Federated Fog vs. PON-Based Non-Federated Fog.

connected to the OLT. On the other hand, there is no PON connectivity among fog cells in the non-federated fog architecture as each cell is connected to a dedicated OLT device, as shown in Fig. 2. The proposed MILP model minimizes the networking and processing power consumption of VM requests. Each VM request consists of a CPU processing demand which is the amount of processing required in Million

Instructions Per Second (MIPS), the traffic demand which is the amount of data required in Gbps, and the RAM workload which is the amount of memory allocated to each VM request in MB.

The following notations are the sets, parameters and variables used in the optimization model

1) *Sets:*

| | |
|---|---|
| $N$ | Set of all nodes in the proposed architecture |
| $N_m$ | Set of all neighbouring nodes to node $m$ in the proposed architecture |
| $ONU_N$ | Set of ONUs in the networking layer, where $ONU_N \in N$ |
| $OLT_N$ | Set of OLTs in the networking layer, where $OLT_N \in N$ |
| $S_P$ | Set of servers in the processing layer, where $S_P \in N$ |
| $VMs$ | Set of all VM request |

2) *Networking Layer Parameters:*

| | |
|---|---|
| $M_{ONU}$ | Maximum power consumption of ONUs. |
| $I_{ONU}$ | Idle power consumption of ONUs. |
| $B_{ONU}$ | Maximum data rate of ONUs. |
| $E_{ONU}$ | Energy per bit of ONU, where $E_{ONU} = \left(\frac{M_{ONU} - I_{ONU}}{B_{ONU}}\right)$. |
| $M_{OLT}$ | Maximum power consumption of OLTs. |
| $I_{OLT}$ | Idle power consumption of OLTs. |
| $B_{OLT}$ | Maximum data rate of OLTs. |
| $E_{OLT}$ | Energy per bit of OLT port, where $E_{OLT} = \left(\frac{M_{OLT} - I_{OLT}}{B_{OLT}}\right)$. |
| $C_{mn}$ | Capacity of the physical link $(m,n)$, where $m, n$ within $N$. |

3) *Processing Layer Parameters:*

| | |
|---|---|
| $M_S$ | Maximum power consumption of servers |
| $I_S$ | Idle power consumption of servers |
| $C_S$ | CPU capacity of the servers |
| $R_S$ | RAM memory capacity of the servers |
| $O_S$ | The proportional power consumption of the servers, where $O_S = M_S - I_S$ |
| $P_d^{ONU}$ | Maximum power consumption of ONUs |
| $D_d^{ONU}$ | Maximum data rate of ONUs |
| $A_S$ | Number of servers permitted to serve one VM request |

4) *Virtual Machines Requests Parameters:*

| | |
|---|---|
| $C_{VMs}$ | CPU demand of VM $s \in VM$ in MIPS. |
| $R_{VMs}$ | RAM memory demand of $VM\ s \in VM$ in GB. |
| $T_{VMs}$ | Traffic Demand of VM $s \in VM$ in Gbps. |

5) *Variables:*

| | |
|---|---|
| $L_{sd}$ | Traffic demand between source node, where $s \in VMs$ and processing device, where $d \in S$. |
| $L_{sd}^{mn}$ | Traffic flow between source node $s \in VMs$ and processing device $d \in S$, traversing node $m \in N$ and $n \in Nm$. |
| $P_{sd}$ | Processing demand between source node $s \in VMs$ and processing device $d \in S$. |
| $R_{sd}$ | RAM memory demand between source node $s \in VMs$ and processing device $d \in S$. |
| $L_m$ | Amount of traffic gathered by node $m \in N$. |
| $B_m$ | $B_m=1$, if the networking node $m \in N$ is activated, otherwise $B_m=0$. |
| $D_m$ | Defined as the AND of two variables $B_m$ and $L_m$, $m \in N$. |
| $B_d$ | $B_d=1$, if the server in the processing layer $d \in S_P$ is activated, otherwise $B_d=0$ |

- The Networking Layer power consumption, ($NET_{PC}$), is composed of:
  a. Power cconsumption of ONUs in the Networking Layer:

$$\sum_{m \in ONU} E_{ONU}\ L_m + \sum_{m \in ONU} I_{ONU}\ B_m \quad (1)$$

  b. Power consumption of OLTs in the Networking Layer

$$\sum_{m \in OLT} E_{OLT}\ L_m + \sum_{m \in OLT} I_{OLT}\ \frac{D_m}{B_{OLT}} \quad (2)$$

- The Processing power consumption, ($P_{PC}$), is composed of :
  a. Power consumption of the servers:

$$\sum_{d \in S} \left( I_S\ B_d + O_S \sum_{s \in VM_s} \frac{(C_{VMs}\ P_{sd})}{C_S} \right) \quad (3)$$

  b. Power consumption of ONUs attached to each server:

$$\frac{P_d^{ONU}}{D_d^{ONU}} \sum_{m \in VM_s} L_m \quad (4)$$

Note that the power consumption of the ONU devices is of two types: 1) the ONUs attached to processing servers which work as transceivers and have an on/off power consumption profile, 2) the ONUs used as CPE. These have a proportional plus an idle power consumption profile [48], [49].

**The MILP model is defined as follows:**

**The objective**: Minimize the total networking and processing power consumption of the proposed architecture:

$$NET_{PC} + P_{PC} \quad (5)$$

*Subject to the following constraints*:

$$\sum_{\substack{n \in N_m \\ m \neq n}} L_{sd}^{mn} - \sum_{\substack{n \in N_m \\ m \neq n}} L_{sd}^{nm} = \begin{cases} L_{sd} & m = s \\ -L_{sd} & m = d \\ 0 & otherwise \end{cases} \quad (6)$$

$$\forall\ s \in VMs, d \in S, \forall m \in N$$

Equation (6) is a traffic flow conservation constraint to ensure that the traffic demand for each VM that enters a node leaves it at same level (except for the source and destination nodes).

$$\sum_{s \in VM_s} C_{VMs} P_{sd} \leq C_s \quad (7)$$

$$\forall\ d \in S$$

Equation (7) ensures that the processing capacity of the VMs' requests does not exceed the processing capacity of the allocated server.

$$\sum_{s \in VM_s} R_{VMs} R_{sd} \leq R_S \quad (8)$$

$$\forall\ d \in S$$

Equation (8) ensures that the memory of the VM requests does not exceed the memory capacity of the allocated server

$$L_m \leq D_d^{ONU} \quad (9)$$

$$\forall\ m \in S$$

Equation (9) ensures that the total traffic aggregated on the server does not exceed the data rate of the ONU.

$$\sum_{s \in VM_s} \sum_{d \in S} L_{sd}^{mn} \leq C_{mn} \quad (10)$$

$$\forall\ n \in N, m \in N_m$$

Equation (10) ensures that the total traffic demand passed through the physical link $m,n$, in all layers does not exceed the capcity of the link.

### III. MILP MODEL RESULT

In this section, we evaluate the efficiency of the proposed model by comparing the Federated-Fog with the Non-Federated Fog over PON as shown in Fig. 2 under three different scenarios of VMs requests (i.e. 10, 15, and 20 VMs) with random uniform distribution for the CPU, memory and traffic demands as shown in Table 1. The VMs' processing requirements are uniformly distributed between (160k MIPS and 280k MIPS) in one cell, and (10k MIPS to 56k MIPS) in the other cells. This results in one fog cell being highly loaded. The VMs memory requirements are uniformly distributed between (100MB and 500MB) [49]. Moreover, The VMs' traffic demands are uniformly distributed between 1 Gbps and 10 Gbps. It is important to note that since the OLT is shared by many users and applications, the idle power consumption of this device should be divided among the applications running at a given point in time, i.e. only a fraction of the maximum idle power consumption is accounted for in the optimization model [48]. The input data used in the MILP model is presented in Table 1.

*Table 1 Input data used in the model*

| | |
|---|---|
| Server's maximum power consumption [50]. | 457 W |
| Server's idle power consumption (66% of Maximum power) [50]. | 301 W |
| Processing capacity of the server [50]. | 280k MIPS |
| Processing capacity of the VMs. | 10k MIPS - 280k MIPS |
| Memory capacity (RAM) of the server [50]. | 16 GB |
| Memory capacity (RAM) of the VMs [49]. | 100 MB - 500 MB |
| OLT Maximum power consumption [51]. | 1940 W |
| OLT idle power consumption (90% of Maximum power. | 1746 W |
| OLT data rate [51]. | 8600 Gbps |
| ONU Maximum power consumption [52]. | 2.5 W |
| ONU idle power consumption (60% of Maximum power) [52]. | 1.5 W |
| ONU data rat e[52]. | 10 Gbps |

| VMs Traffic Demands [49]. | 1 Gbps – 5 Gbps |
| Capacity of Optical physical link [53]. | 32 wavelengths per fibre at 40 Gbps per wavelength |

We have evaluated the server utilization of both architectures by comparing the number of VMs processed versus the number of VMs blocked. Figure 3 shows that with the Non-Federated Fog approach, 3 VMs are blocked in total due to the limited local processing resources available to each fog cell. Fig. 4, on the other hand, shows that each cell in the Federated

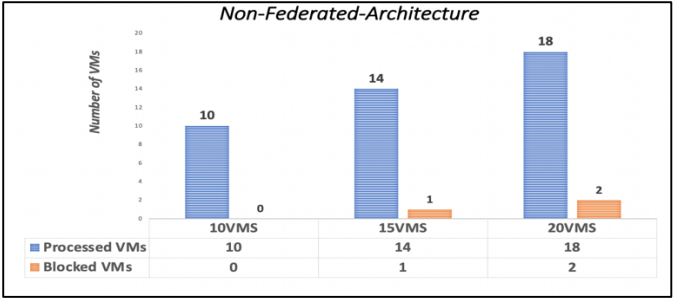

*Figure 3 Number of processed versus blocked VMs in the Non-Federated Fog.*

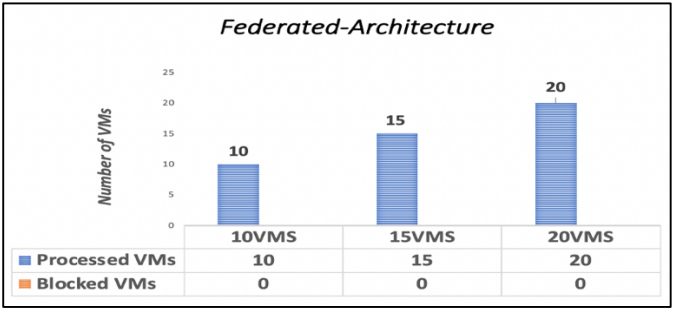

*Figure 4 Number of processed vs. blocked VMs in the Federated Fog*

Fog architecture is capable of borrowing data processing from neighbouring fog cells, hence, no VMs are blocked with this approach.

Moreover, we have compared the total power consumption of both approaches. As can be seen in Fig. 5, the power savings in the Federated Fog architecture for processing 10, 15 and 20 VMs is up to 26%, 7% and 2%, respectively. The processing power consumption required to accommodate the assigned VMs and the number of activated servers, as shown in Fig. 6 and Fig. 7 were the main contributing factors in the power savings. In the Federated Fog architecture, the model is able to pack servers and hence use fewer processing servers due to the federation facilitated by the PON connectivity. At 10 VMs, both approaches have processed the entire VMs demands, as shown in Fig. 3 and Fig. 4. However, as can be observed in Fig. 7, the difference is in the power consumed for processing in the VMs, as with the Non-Federated Fog, a higher number of servers were utilized as opposed to the Federated Fog approach. When considering the placement of 20 VMs, the Non-Federated Fog has used 7 servers to accommodate 18 VMs out of 20 VMs while the Federated Fog has used 6 servers to accommodate all of the VMs, as shown in Figure 7. Obviously, due to the PON devices used in the networking layer, both architectures are highly energy efficient in networking power consumption and the processing power consumption has the greatest weight in terms of where to process the VMs, as shown in Figure 6. An interesting direction for future work is to consider different levels of resource efficiency in the networking and processing layers as there can be intrinsic trade-offs in this regard.

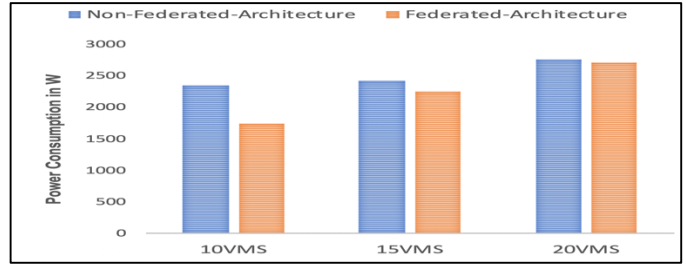

*Figure 5 Total power consumption of both fog architectures.*

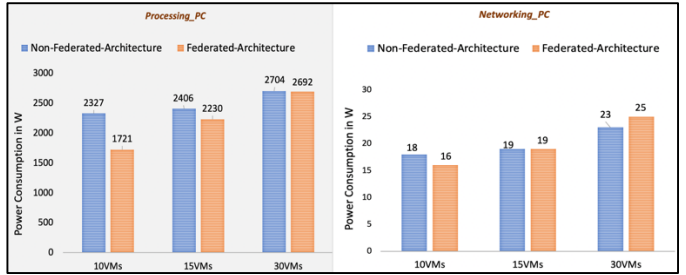

*Figure 6 Networking power consumption versus processing power consumption in both fog architectures.*

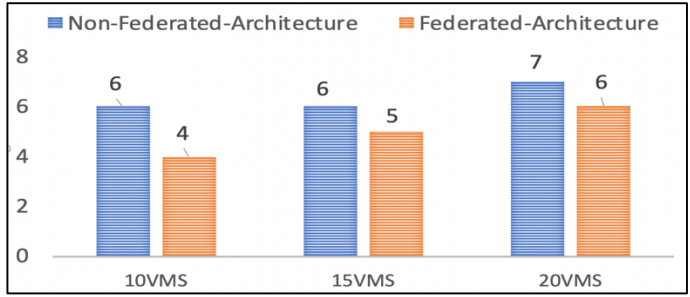

*Figure 7 Server utilization in both fog architectures.*

IV. CONCLUSIONS

In this paper, we proposed a federated fog computing architecture where a number of fog computing units in the access network are connected through a PON. This enables providing higher processing capacity in the fog layer that can process more data closer to end-users. We compared the Federated Fog approach to the Non-Federated Fog over a PON access network. The results showed that through the Federated approach, a total power saving of up to 26% can be achieved compared to the Non-Federated Fog approach. Moreover, the total number of VMs blocked can be reduced. Future work includes extending the optimization model to consider a weighted objective function that incorporates delay and power. Also, consideration will be given to mobility-aware workload assignment which is an interesting future research direction.

ACKNOWLEDGMENTS

The authors would like to acknowledge funding from the Engineering and Physical Sciences Research Council (EPSRC) INTERNET (EP/H040536/1), STAR (EP/K016873/1) and TOWS (EP/S016570/1) projects. The first author would like to acknowledge the Government of Saudi Arabia and JAZAN University for funding his PhD scholarship.